# AI climate trajectories shaped by growth, power choices and decarbonization

Xiaoyu Xie[1], John Bistline[2,3], Ming Xu[1]*, Sangwon Suh[1]*
[1] School of Environment, Tsinghua University, Beijing 100084, China
[2] Watershed Technology Inc., San Francisco, CA 94103, USA
[3] MIT Sloan School of Management, Cambridge, MA 02139, USA
* Corresponding authors: xu-ming@tsinghua.edu.cn; sangwon@tsinghua.edu.cn

Abstract: Artificial intelligence (AI) is emerging as a major source of electricity demand, yet its operational greenhouse gas (GHG) emissions remain difficult to resolve geographically. We develop a facility-level global inventory of AI-oriented data centers and extend it to 2030 using documented expansion plans and grid-policy scenarios. Under Current Policies, emissions increase from 20.4 $MtCO_2e$ in 2024 to 68.7 $MtCO_2e$ in 2030, with parameter-side Monte Carlo 5th–95th percentile ranges of 19.3–21.0 and 60.8–73.9 $MtCO_2e$, respectively. Only six geographies, representing less than 3% of projected capacity, avoid rising emissions. Grid decarbonization consistent with net zero by 2050 lowers projected 2030 emissions to 48.3 $MtCO_2e$, but fossil-based behind-the-meter generation erodes this benefit. Under that scenario, the U.S. behind-the-meter emissions penalty exceeds the combined AI-data-center emissions outside the United States. These findings support coordinated capacity planning, faster access to low-carbon electricity and limits on routine fossil-based behind-the-meter supply.

## 1 Introduction

A race between two opposing forces is shaping the climate trajectories of Artificial Intelligence (AI): rapid growth in electricity demand by AI data centers and declining grid emissions intensity. Global electricity demand from AI-oriented data centers is projected to more than quadruple between 2024 and 2030[1]. At the same time, the global electricity grids have been decarbonizing over the past decade, with the rate accelerating in recent years amid record renewable-power deployment[2].

However, despite growing estimates of AI's electricity demand and decarbonizing grids, quantifying the greenhouse gas (GHG) emissions from AI-oriented data center operations and projecting their trajectories remain challenging. This challenge is threefold. First, current global estimates of GHG emissions inventory of AI-oriented data centers are inferred from proxies, such as accelerated-server shipments, which reveal neither where infrastructure is deployed nor how intensive the GHG emissions of the electricity it consumes—both of which are crucial in understanding the GHG inventory of AI-oriented data centers[3,4,5,6]. For example, GHG emissions intensities of electricity range from about 7 $gCO_2e\ kWh^{-1}$ in Sweden and 43 $gCO_2e\ kWh^{-1}$ in France to about 554 $gCO_2e\ kWh^{-1}$ in Poland[7]. The range remains substantial at subnational scale for larger countries; subregional grid emissions intensities in the United States, for example, span from 110 to 514 $gCO_2e\ kWh^{-1}$ [8]. Second, not only grid carbon intensity but also the pace of decarbonization remain highly heterogeneous across regions. Over the last decade, the carbon intensity of electricity in the United Kingdom fell by ~70%, from 419 to 124 $gCO_2\ kWh^{-1}$, as coal was phased out and renewable generation more than doubled[9] , whereas India's weighted-average grid emission factor declined by ~9%, from 779 $gCO_2$ $kWh^{-1}$ to 710 $gCO_2$ $kWh^{-1}$ during the same period[10]. Third, grid congestion and long interconnection queues have made behind-the-meter (BTM) generation an emerging trend for new data centers[1,11,12]. New transmission lines can take four to eight years to build in advanced economies, whereas natural-gas turbines can be installed within weeks[10,11]. Such BTM gas generation can accelerate AI data center deployment in rapidly expanding regions, but adds a fossil carbon burden that bypasses the utility grid and its decarbonization trajectories.

Therefore, understanding GHG emissions from AI-oriented data centers and how they evolve requires location-specific data on current operations, expansion characteristics, power source choices, grid GHG intensities and grid-decarbonization trajectories [13]. However, acquiring such facility-level, location-specific data is not straightforward[14,15,16]. First, published records rarely specify whether a data center is used mainly to render AI-related services or other services such as cloud computing, streaming and web hosting[17,18,19,20]. Second, plans for data-center capacity expansion and power-source choices, including BTM generation, are fragmented across regional and local planning documents, permit applications and corporate announcements, making systematic data collection and harmonization labor-intensive.

To fill this gap, we compiled a global inventory of 9,946 data centers by systematically identifying, reconciling, and validating information from more than 27,000 pieces of fragmented public evidence, and classified AI-oriented data centers using multiple infrastructure signals, following the framework described in ref. [21] (Supplementary Note 6). AI-oriented facilities differ from conventional data centers in a number of key characteristics[19,22,23,24,25] because they are organized around dense deployments of accelerated servers—typically GPU-rich nodes coupled with higher-power CPUs, larger memory footprints, and high-bandwidth interconnects for AI training and inference—which drive substantially higher and more sustained facility-level power densities[1,4,26]. Therefore, AI-oriented data centers can be distinguished from the rest through multiple infrastructure signals. For example, high-density AI workloads often require greater electrical capacity[27,28,29,30], more intensive thermal management[23,31,32], and facility adaptations such as liquid-cooling systems[33,34], reinforced structures for heavy racks and cooling equipment, or other forms of high-density build-out[35] (see Supplementary Table 12 ). We then integrated data on AI-oriented data center locations with location-specific GHG emissions intensities of electricity and facility-specific on-site power generation to derive location-specific GHG inventory of AI-oriented data centers following the Greenhouse Protocol Corporate standard (see Section 2 for details)[36]. We then combine documented expansion plans with geography-specific grid decarbonization trajectories under three representative scenarios, including Current Policies, Nationally Determined Contributions (NDCs) and Net Zero by 2050 scenarios, under three integrated assessment models including GCAM, MESSAGEix-GLOBIOM and REMIND-MAgPIE[37]; for the United States, given its outsized share in the global AI infrastructure, we further differentiate these trajectories at the state level using scenario-consistent projections (see Section 2). We account for documented routine BTM supply by blending grid and source-specific direct emission factors according to the share of facility electricity supplied. This framework enables geographically explicit estimates of AI-oriented data center electricity use, power system choice, and annual greenhouse gas emissions inventory.

## 2 Materials and methods

### 2.1 Data sources and study design

The global inventory was constructed by facility-level candidate discovery with site-specific public evidence. The corpus comprised 14,939 unique webpages from operators, governments, utilities, grid records, certification materials, public reporting and geospatial sources[21]. Commercial directories supported discovery and cross-validation, while public sources supplied traceable facility attributes. Records were linked through operator, site and geographic identifiers and harmonized across identity, location, development stage and technical characteristics. Identity resolution, campus consolidation and consistency checks produced 9,946 facilities worldwide.

Identification of AI-oriented facilities relied on a separate layer of site-specific evidence. We screened 2,280 public evidence records linked to 451 AI-relevant candidate facility observations, of which 181 met the site-level criteria for AI-specific or AI-optimized deployment[21]. For facilities lacking sufficiently specific public evidence of AI deployment, we did not infer the absence of AI activity. This included general-computing facilities for which no site-level AI evidence was identified. Such facilities were therefore retained as unresolved rather than classified as confirmed non-AI sites. This positive-unlabeled design treats missing public evidence as non-observability rather than evidence against AI orientation.

### 2.2 Evidence-linked tiered screening

The method of identifying AI-oriented data centers that we employed in this study uses a hierarchical framework based on a gradient-boosted decision tree (GBDT) trained on facility-level public disclosures on infrastructure signals.[21] Given that publicly available disclosure of AI-oriented data centers are sparse, while unlabeled data centers are not necessarily non-AI data centers, the method employs a positive-unlabeled learning (PU-learning) framework under incomplete observability.[38,39,40,41] The method preserves ranking performance after interaction augmentation and produces interaction splits that can be directly translated into physically interpretable threshold rules for deployment.[21] In this framework, an AI-oriented data center is defined as "a facility or consolidated campus for which either (1) site-level public evidence establishes AI-specific hardware, workloads or purpose-built infrastructure, or (2) recorded infrastructure characteristics satisfy prespecified GBDT-derived criteria learned from facilities with such evidence".[21] The definition includes both dedicated AI facilities and mixed-use facilities with supported AI deployment or AI-oriented infrastructure.

The model used power capacity, rack density, cooling architecture, liquid-cooling status, PUE and facility scale, represented through continuous, binned and hybrid encodings. Learned interactions were compressed into threshold rules that combined distinct infrastructure signal groups and were evaluated through recall, placebo, missingness and manual checks. For rule $r$ in company-holdout test split $t$, let $a_{r,t}$ denote the observed anchor prevalence among facilities selected by the rule and $a_t$ the anchor prevalence in the full test split. Rule enrichment[41,42] was

$$e_{r,t} = \frac{a_{r,t}}{a_t}.$$

Values greater than one indicate above-baseline prevalence of observed AI evidence in unseen-company samples. The model therefore served as an evidence-enrichment screen rather than a deterministic AI/non-AI classifier.

Direct site-bound evidence and validated infrastructure-rule support were combined into ordinal evidence tiers. Confirmed facilities received full inclusion, while facilities near the identification boundary received evidence-conditioned weights. For perturbation-eligible boundary-tier facility $i$, the central expected inclusion weight was

$$w_i = \frac{1}{B}\sum_{b=1}^{B} z_i^{(b)}, \qquad B = 5000.$$

Here $z_i^{(b)}$ is a binary indicator equal to one when facility $i$ is reassigned to the AI core in perturbation $b$ under the medium-noise profile and zero otherwise. Confirmed facilities were assigned $w_i = 1$; the complete category-to-weight mapping, planning-stage rules and 2024 observation-year adjustment follow the approach described in ref. [21] (Supplementary Note 8). The resulting $w_i$ is an evidence-side expected inclusion weight. It is neither a fully calibrated probability that a facility is AI-oriented nor an estimate of the fraction of facility electricity consumed by AI workloads (Supplementary Note 8).

### 2.3 Facility power and annual electricity

Next, we estimate facility power in AI data centers[43,44]. Facility power was assigned through a fixed nine-route priority hierarchy rather than by selecting the maximum across available proxies. The hierarchy proceeds from reported IT or validated site power through rack-, space-, building-, partial-rack- and accelerator-based reconstruction, with an ambiguous direct-power route used only when no higher-priority route is available. Route order, plausibility checks and imputation procedures follow the method described in ref. [21].

Specifically, the directly observed rack, space and accelerator routes were calculated as

$$P^{\mathrm{IT}}_{\mathrm{rack},i} = \frac{N_{\mathrm{rack},i}\, P^{\mathrm{typ}}_{\mathrm{rack},i}}{1000}.$$

$$P^{\mathrm{IT}}_{\mathrm{space},i} = \frac{A_{\mathrm{space},i}\, 10.7639\, \rho_{\mathrm{area},i}}{10^6}.$$

$$P^{\mathrm{IT}}_{\mathrm{card},i} = \frac{N_{\mathrm{accel},i}\, k^{\mathrm{IT}}_{\mathrm{card},i}}{1000}.$$

where $N_{\mathrm{rack},i}$ is rack count, $P^{\mathrm{typ}}_{\mathrm{rack},i}$ is typical rack power in kW, $A_{\mathrm{space},i}$ is white-space area in m$^2$, $\rho_{\mathrm{area},i}$ is area power density in W ft$^{-2}$, $N_{\mathrm{accel},i}$ is site-level accelerator count and $k^{\mathrm{IT}}_{\mathrm{card},i}$ is effective IT power per card in kW. Each expression yields an IT-load estimate in MW; imputed variants use the same transformations with missing terms drawn from matched strata. The accelerator route is used only when higher-priority reported IT power, validated site power, rack, space, building and partial-rack routes are unavailable.

For facility $i$ in year $y$, annual electricity use was calculated as

$$E_{i,y} = 8760\, P_i\, p^{\mathrm{realize}}_{i,y}\, \mathrm{ramp}_{i,y}\, LF_{i,y}\, \mathrm{PUE}^{I_i}_{i,y},$$

where $P_i$ is the selected power estimate in MW, $p^{\mathrm{realize}}_{i,y}$ is the development-stage realization factor, $LF_{i,y}$ is the annual load factor and $I_i = 1$ for IT-load routes and $I_i = 0$ for facility- or nameplate-load routes. Rack-, space-, accelerator- and IT-based imputation routes therefore receive a PUE multiplier, whereas validated site power and the ambiguous direct-power route do not.

Commissioning followed

$$\mathrm{age}_{i,y} = y - t_{0,i},$$

and

$$\mathrm{ramp}_{i,y} = \begin{cases} 0, & \mathrm{age}_{i,y} < 0, \\ \min\left(1, \dfrac{\mathrm{age}_{i,y} + 1}{R_i}\right), & \mathrm{age}_{i,y} \geq 0, \end{cases}$$

where $t_{0,i}$ is the commissioning year and $R_i$ is the archetype-specific ramp duration. The $+1$ convention assigns $1/R_i$ of annual ramped load in the commissioning year. Evidence-weighted electricity use in region $r$ was

$$E_{r,y}^{\mathrm{AI}} = \sum_{i:g(i)=r} w_i\, E_{i,y},$$

where $g(i)$ denotes the grid region assigned to facility $i$.

## 2.4 Operational electricity-related GHG accounting

We limit the facility-level accounting boundary within Scope 1 and Scope 2 inventory emissions (as opposed to life-cycle emissions) from onsite generation and purchased electricity following the GHG Protocol Corporate Standard and Scope 2 Guidance[36]. We estimated facility-level operational GHG emissions from electricity use using a physical, location-based accounting framework. The boundary covers direct emissions from electricity generation. Grid electricity was assigned annual location-based emission factors. For documented physically dedicated behind-the-meter supply, grid and source-specific direct emission factors were blended according to the supplied electricity share. Dedicated generation corresponds to Scope 1 only where the reporting entity owns or controls the generating asset; because ownership and control were not consistently observable, we report emissions based on facility-specific electricity use rather than company-specific Scope 1 and Scope 2 inventories. The boundary excludes upstream fuel supply, embodied and life-cycle emissions, and reductions claimed solely through market-based contractual instruments.

Grid-supplied electricity received the annual location-based factor of the facility's electricity region. A BTM source replaced the grid factor only when evidence established dedicated physical supply for routine facility demand and a supportable coverage share. Grid-connected power-purchase agreements, nearby or contractually co-located generation, backup-only assets and small auxiliary sources retained the grid factor when routine load remained grid supplied. Partial direct supply was blended according to the documented or supportable coverage fraction. The current implementation assigns at most one dominant BTM technology to each facility; mixed systems are mapped to the dominant class.

The effective emission factor, $EF_{\mathrm{g_i,y,s,m}}^{\mathrm{grid}}$ , is the location-based grid emission factor assigned to facility $i$ through region $g_{\mathrm{i}}$, $EF_{\mathrm{i,y}}^{\mathrm{BTM}}$ is the direct emission factor of its qualifying behind-the-meter source, and $EF_{\mathrm{i,y,s,m}}^{\mathrm{eff}}$ is the resulting blended facility factor.

$$EF_{i,y,s,m}^{\mathrm{eff}} = \left(1 - b_{i,y}\right) EF_{g(i),y,s,m}^{\mathrm{grid}} + b_{i,y} EF_{i,y}^{\mathrm{BTM}},$$

where $b_{i,y}$ is the share of facility electricity physically supplied by the qualifying dominant BTM source, $s$ denotes the policy scenario and $m$ denotes the integrated assessment model. Facility-weighted regional emissions were

$$G_{r,y,s,m}^{\mathrm{AI}} = \sum_{i:g(i)=r} w_i \, E_{i,y} EF_{i,y,s,m}^{\mathrm{eff}},$$

with unit conversion from MWh multiplied by $gCO_2e$ $kWh^{-1}$ to $MtCO_2e$ applied after aggregation. BTM classes distinguish simple-cycle gas, combined-cycle gas, reciprocating gas engines, natural-gas solid-oxide fuel cells, wind, directly supplied nuclear power and hydrogen fuel cells. Source-specific direct factors and their boundaries are reported in the Supplementary Methods.

Electricity use was estimated from the tiered facility inventory using the hierarchical physical load model described above, prioritizing directly reported site power where available and otherwise selecting an infrastructure-based estimation route according to data availability and plausibility. Facility electricity use was matched to subnational grid emission factors where available and to country-level factors otherwise. For 2024, we assigned state-level eGRID factors in the United States[45], consumption-based factors for European countries constructed from JRC national production factors[46] allocated through annual partner-trade proportional sharing of Eurostat bilateral trade[47], and direct, flow-traced consumption factors from Electricity Maps[48] for all other countries; the China national factor is from the official 2024 $CO_2$ value and applies a non-$CO_2$ uplift derived from the observed generation mix under IPCC AR6 $GWP_{100}$.

**2.5 Policy-conditioned grid scenarios for 2030**

The 2030 analysis comprises policy-conditioned projections under the Current Policies, NDCs and Net Zero 2050 scenarios[37]. All scenarios share the same facility inclusion, AI-oriented classification, development plans, power reconstruction, realization, utilization, electricity-demand trajectory and documented behind-the-meter configurations. Grid emission-factor trajectories provide the sole scenario-varying input, isolating the emissions effect of alternative power-sector pathways under a common data-center expansion pathway. National trajectories were derived separately from GCAM 6.0, MESSAGEix-GLOBIOM 2.0-M-R12-NGFS and REMIND-MAgPIE 3.3–4.8. Reported central estimates use the cross-model median, with model-specific results retained to characterize structural variation. For geography $g$, policy scenario $s$ and model $m$, the 2024 location-specific factor was scaled by the corresponding IAM trajectory:

$$EF_{g,2030,s,m}^{\mathrm{grid}} = EF_{g,2024}^{\mathrm{grid}} R_{g,s,m}.$$

Here, $R_{\mathrm{g,s,m}}$ is the dimensionless ratio of the IAM-modelled grid emission intensity in 2030 to that in 2024 for geography $g$ under policy scenario $s$ in model $m$.

The central factor was calculated separately for each geography and policy scenario as

$$\widetilde{EF}_{g,2030,s}^{\mathrm{grid}} = \mathrm{median}_{m\in\{\mathrm{GCAM,\,MESSAGEix,\,REMIND}\}} \left( EF_{g,2030,s,m}^{\mathrm{grid}} \right).$$

Main-text 2030 estimates use the region-by-region three-model median under Current Policies. NDCs and Net Zero 2050 provide policy sensitivities[49], and all nine model-scenario values are reported in Supplementary Table 22. In the United States, Cambium supplied the state-level spatial pattern, normalized to reproduce the corresponding IAM national factor[50]. Facility and electricity-demand assumptions were held constant across policy scenarios; only the grid emission-factor pathway varied.

These projections are conditional on documented BTM configurations and do not dynamically model subsequent transitions to grid supply. The inclusion of these configurations reflects the use of off-grid power to circumvent multi-year interconnection delays[51]. IAM-specific variation is reported as model detail and is not included in the headline uncertainty interval.

### 2.6 Parameter-side Monte Carlo uncertainty

Uncertainty in the electricity model was propagated using a parameter-side Monte Carlo framework that held route structure fixed while sampling approved parameter families, including realization fractions, commissioning years, ramp duration, load factor, imputed PUE, and imputation-dependent power terms. Directly observed power routes were held fixed, whereas imputation-dependent routes propagated uncertainty through stratified donor pools or bounded parametric distributions. Stratification grouped facilities by comparable project characteristics, including development stage, facility type, route class, and size-related proxies, so that sampled parameters reflected heterogeneity within similar infrastructure classes rather than pooled global variation. This design separated infrastructure-side uncertainty from downstream AI-identification uncertainty. AI-attributable electricity and GHG totals were calculated only after facility-level electricity and location-based emissions had been estimated, by applying independently derived facility-level AI evidence weights. Route definitions, baseline parameter selection, imputation logic and stratification follow the methods described in ref. [21]. The assumptions and Monte Carlo settings used for the 2030 projections are provided in Supplementary Notes 9 and 12.

The default isolated Monte Carlo setting used 5,000 iterations and a fixed random seed. Realization fractions, commissioning years, ramp duration, and load factors were sampled from strata-specific distributions, whereas route-specific power terms and imputed PUE values were drawn from matched donor-pool bootstrap distributions or, where donor support was insufficient, from bounded PERT distributions with hierarchical backoff across related strata. Emission factors were held fixed in this layer so that the uncertainty envelope captured infrastructure reconstruction, stage realization, efficiency, and utilization uncertainty (see Supplementary Note 12).

## 3 Results

### 3.1 Spatial concentration of capacity

AI-oriented facilities identified are heavily concentrated at every geographic scale (Fig. 1; see Section 2 and ref. [21] for identification methods and validation; Supplementary Note 6 and Supplementary Table 12). In 2024, identified AI-oriented capacity corresponds to 13.8 GW globally, of which North America accounts for 9.3 GW and the United States alone for 9.2 GW, or 67% of the global total, with Texas (1.9 GW), Virginia (1.8 GW) and Tennessee (1.1 GW) contributing the most (Fig. 1C).

According to the publicly disclosed capacity expansion plans, AI-oriented capacity could reach 48.5 GW of power capacity globally by 2030 (Fig. 1B). Although comprising only 391 facilities (4% of identified facilities), they account for 29% of observed global data center power capacity by then due to their larger size (Fig. 1B). Concentration persists with the United States accounting for 35.3 GW, or 73% of the global total, followed by China (2.5 GW) and the United Kingdom (1.4 GW). Together, these three countries account for 81% of global AI-oriented capacity, and the ten largest national markets account for 93%. Within the United States, capacity is concentrated in Texas (12.1 GW), Louisiana (3.8 GW),

Pennsylvania (3.6 GW) and Virginia (3.0 GW). At the facility-level, the four largest campuses (A–D in Fig. 1C) together carry 10.1 GW of expected realized load (Fig. 1C).

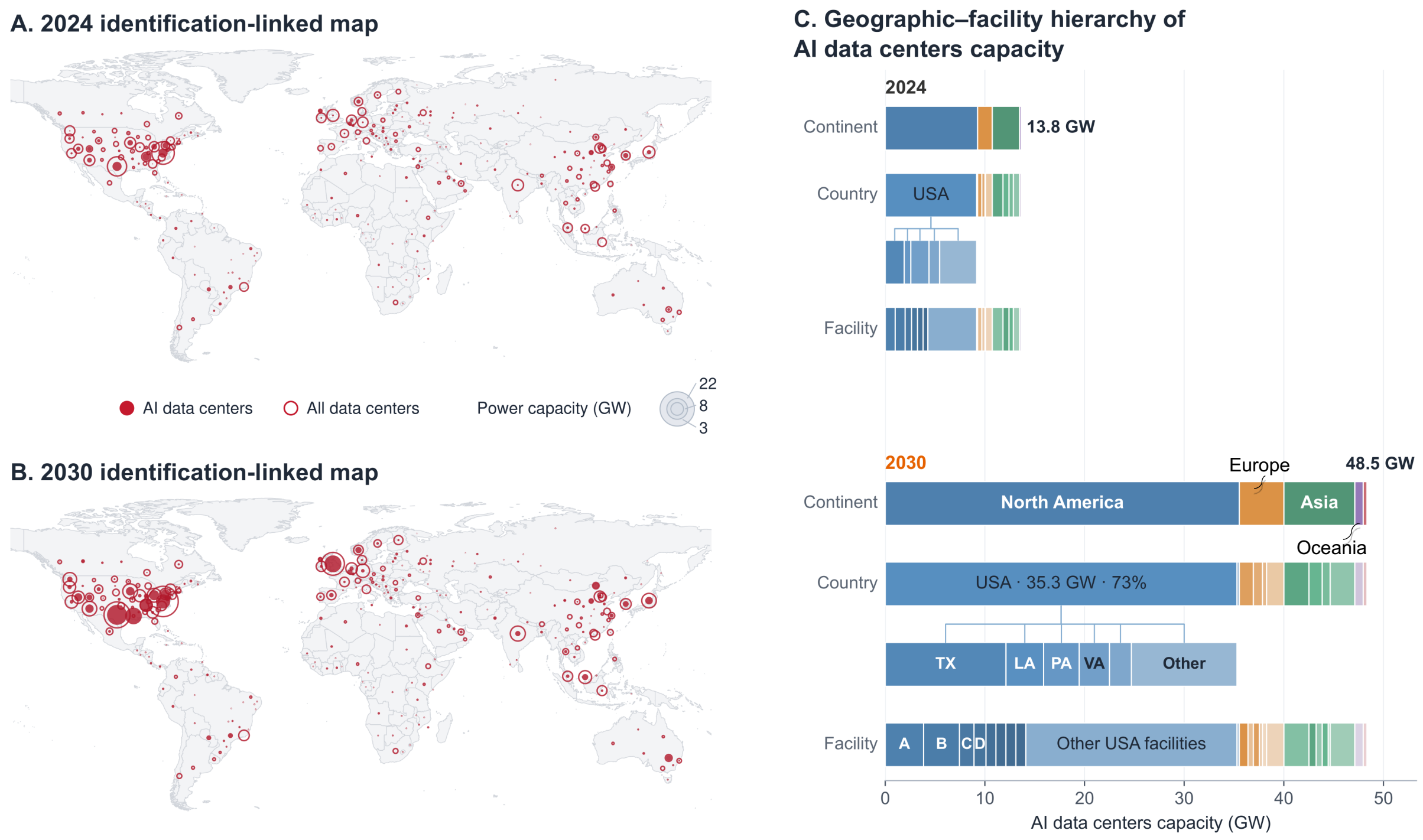

Fig. 1. Identification-linked geography and multi-scale concentration of AI-oriented data centers in 2024 and 2030. (A) Identification-linked geography of AI-oriented (filled circles) and other (open circles) data centers in 2024; circle area denotes power capacity. (B) The same geography for 2030, including capacity under construction and in planning. (C) Geographic–facility hierarchy of expected realized AI data center capacity in 2024 (top) and 2030 (bottom), decomposed from continent to country, U.S. state and individual facility. Letters mark the largest single campuses: A, Project Matador Campus (Fermi America; Texas); B, Hyperion Campus (Meta; Louisiana); C, Wildcat Ridge AI Data Center Campus (operator undisclosed; Pennsylvania); D, Stargate Abilene Campus (Crusoe; Texas);

## 3.2 Location-specific emissions in 2024 and 2030

Location-specific GHG attribution reveals that AI electricity demand and emissions already diverge in 2024, and this divergence widens by 2030. In 2024, AI-oriented facilities consume 57.2 TWh (p5-p95 range, 54.8–59.1 TWh; see Supplementary Note 3, Supplementary Fig. 2) and emit 20.4 $MtCO_2e$ (Monte Carlo p5-p95 range, 19.3-21.0 $MtCO_2e$), with the United States contributing 38.0 TWh and 14.2 $MtCO_2e$, or 66% of AI electricity demand and 70% of AI-related emissions, respectively (Fig. 2A) (see Supplementary Note 9 for the details on capacity factors). The United States, China and South Korea together account for 45.5 TWh and 17.8 $MtCO_2e$, concentrating the emerging footprint within a narrow geographic core.

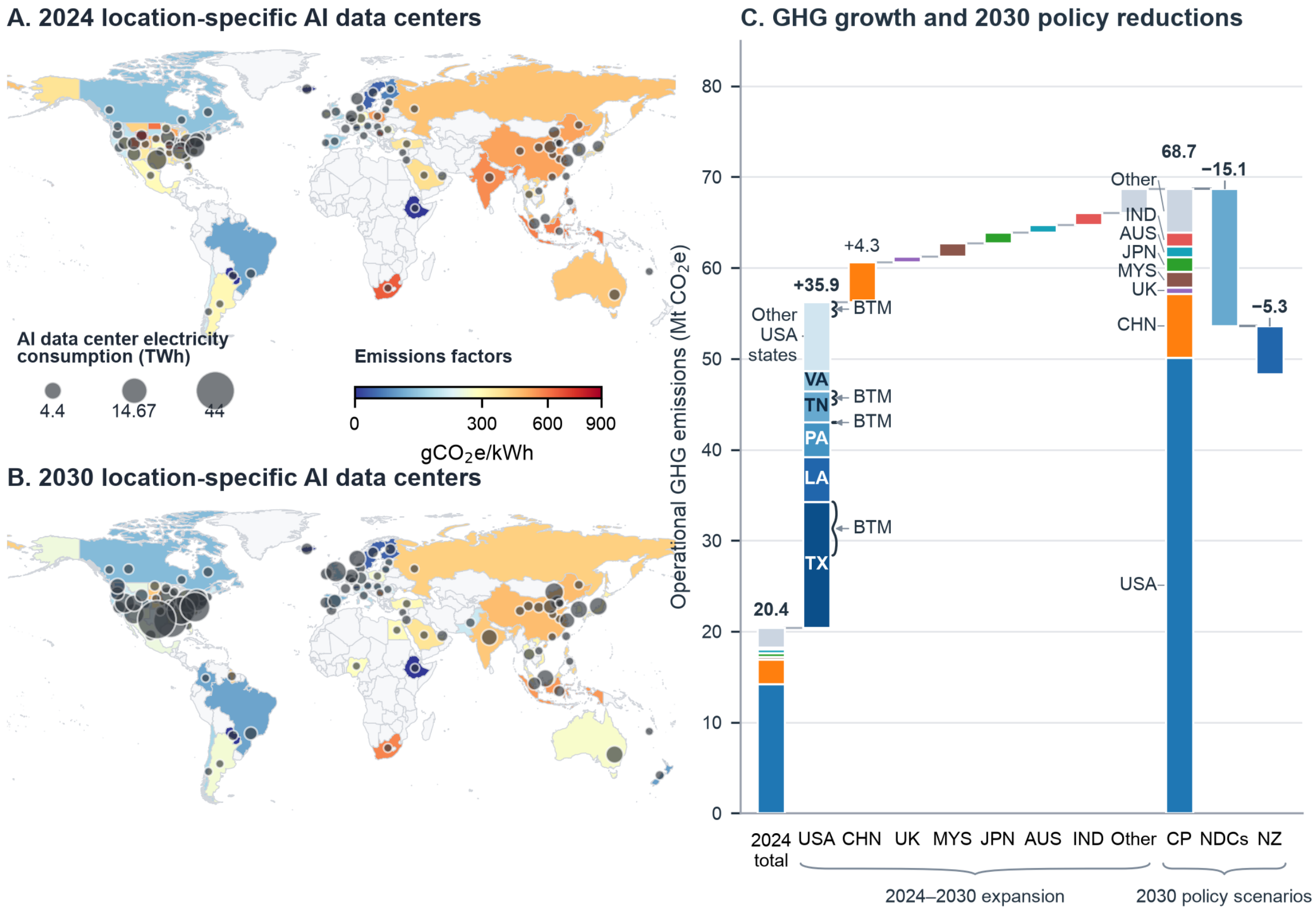

Fig. 2. Location-specific attribution reshapes the AI electricity and GHG footprint in 2024 and 2030. (A) Location-specific distribution of AI data center electricity use in 2024; circle area denotes national AI electricity consumption and map shading gives the location-specific emission factor. (B) The same distribution for 2030 under the Current Policies scenario. (C) Waterfall decomposition of AI-related GHG emissions growth from 20.4 $MtCO_2e$ in 2024 to 68.7 $MtCO_2e$ in 2030 under the Current Policies scenario (CP), resolved by country, with the U.S. increment further resolved by state and by documented BTM supply, with brace heights denoting the BTM contribution to emissions growth and the alternative 2030 outcomes under the NDCs scenario and the Net Zero 2050 scenario (NZ).

Under the Current Policies scenario, global AI electricity demand reaches approximately 250 TWh and expands across power systems with sharply different decarbonization rates by 2030 (Fig. 2B). The United States remains the dominant locus of AI expansion, with AI electricity consumption rising to 177.3 TWh and emissions to 50.1 $MtCO_2e$ with 35.9 $MtCO_2e$ additional emissions (implied GHG intensity of 283 $gCO_2e/kWh$) (Fig. 2 A, C). In the United Kingdom, rapid expansion of AI-oriented data center capacity raises electricity use to 8.1 TWh and GHG emissions only to 0.7 $MtCO_2e$ by 2030 (86 $gCO_2e/kWh$), while China reaches 15.2 TWh—about twice the UK total—and emits 7.0 $MtCO_2e$ (461 $gCO_2e/kWh$) (Fig. 2 B, C).

Taken together, GHG emissions from AI-oriented data center operation are expected to reach 68.7 $MtCO_2e$ (Monte Carlo p5-p95 range, 60.8-73.9 $MtCO_2e$) under Current Policies scenario in 2030 (Fig. 2C), which is a 3.4-fold increase from 2024 level. From this, the emissions fall by 15.1 $MtCO_2e$ under NDC scenario and by another 5.3 $MtCO_2e$ under Net Zero 2050 scenario to reach 48.3 $MtCO_2e$ (Monte Carlo p5-p95 range, 41.6-53.9 $MtCO_2e$; Fig. 2C). Nevertheless, even the Net Zero 2050 case leaves emissions at about 2.4 times the 2024 level.

### 3.3 Grid decarbonization and behind-the-meter generation

Our results show that GHG emissions from AI-oriented data-center operation can be meaningfully moderated only under aggressive grid-decarbonization pathways. In the Netherlands, for example, AI electricity demand increases by 67% while emissions rise by only 19%, as projected grid decarbonization under current policies absorbs most of the additional demand. Similarly, Singapore's AI-related electricity demand increases by 58% while emissions rise by 27%. In contrast, Malaysia and India exhibit near proportional increase in GHG emissions alongside AI-oriented data center capacity expansion because grid carbon intensity remains high; AI electricity demand rises 8.1-fold in Malaysia and 9.7-fold in India, while AI-related emissions increase 6.4-fold and 7.5-fold, respectively (Supplementary Note 13, Supplementary Fig. 3 and Supplementary Table 19).

The deviation of the country- and state-specific trajectories from the iso-emission contours (curves from the upper left to the lower right) shows how far projected emissions trajectories depart from a constant-emissions pathway. Among the major hosts of AI-oriented data centers, South Korea's trajectory remains most closely aligned with the iso-emission contour (Fig. 3A). The trajectory of the United States deviates substantially from the iso-emission line reaching 50.1 Mt$CO_2$e and accounting for 73% of the global total, as its electricity use rises 4.7-fold and its implied emission factor, which accounts for the planned BTM emissions intensity, falls by one quarter. The United Kingdom, one of the least GHG-intensive hosts, converts a 28-fold capacity expansion measured in power consumption into a 21-fold emissions increase. Emissions also rise 2.6-fold in China, 3.8-fold in Japan and 7.5-fold in India, where grid emission factors remain near or above the 2024 IEA global average[52].

The dominant U.S. total reflects two state-level patterns: exceptionally rapid build-out and the proliferation of behind-the-meter generation that weakens the benefit of grid decarbonization (Fig. 3B). Texas illustrates the scale of build-out rate clearly: electricity demand rises from 5.8 to 50.8 TWh and emissions from 2.0 to 16.0 Mt$CO_2$e, exceeding the 2030 total of every country except the United States. Louisiana's 2030 share is entirely driven by planned capacity expansion, reaching 18.0 TWh and 4.9 Mt$CO_2$e from near zero. The wide use of behind-the-meter (BTM) generation creates a second pattern: documented on-site natural-gas generation behind the meter raises Tennessee's implied emissions intensity from the grid average of 330 to 512 g $CO_2$e $kWh^{-1}$ and keeps the implied emissions factors of Texas and Mississippi nearly flat even though their state grids continue to decarbonize (Fig. 3B). Wyoming, Oregon and Utah form the opposite pattern, where moderate build-out rate allows grid-decarbonization to reduce absolute emissions (Supplementary Note 13 and Supplementary Table 19).

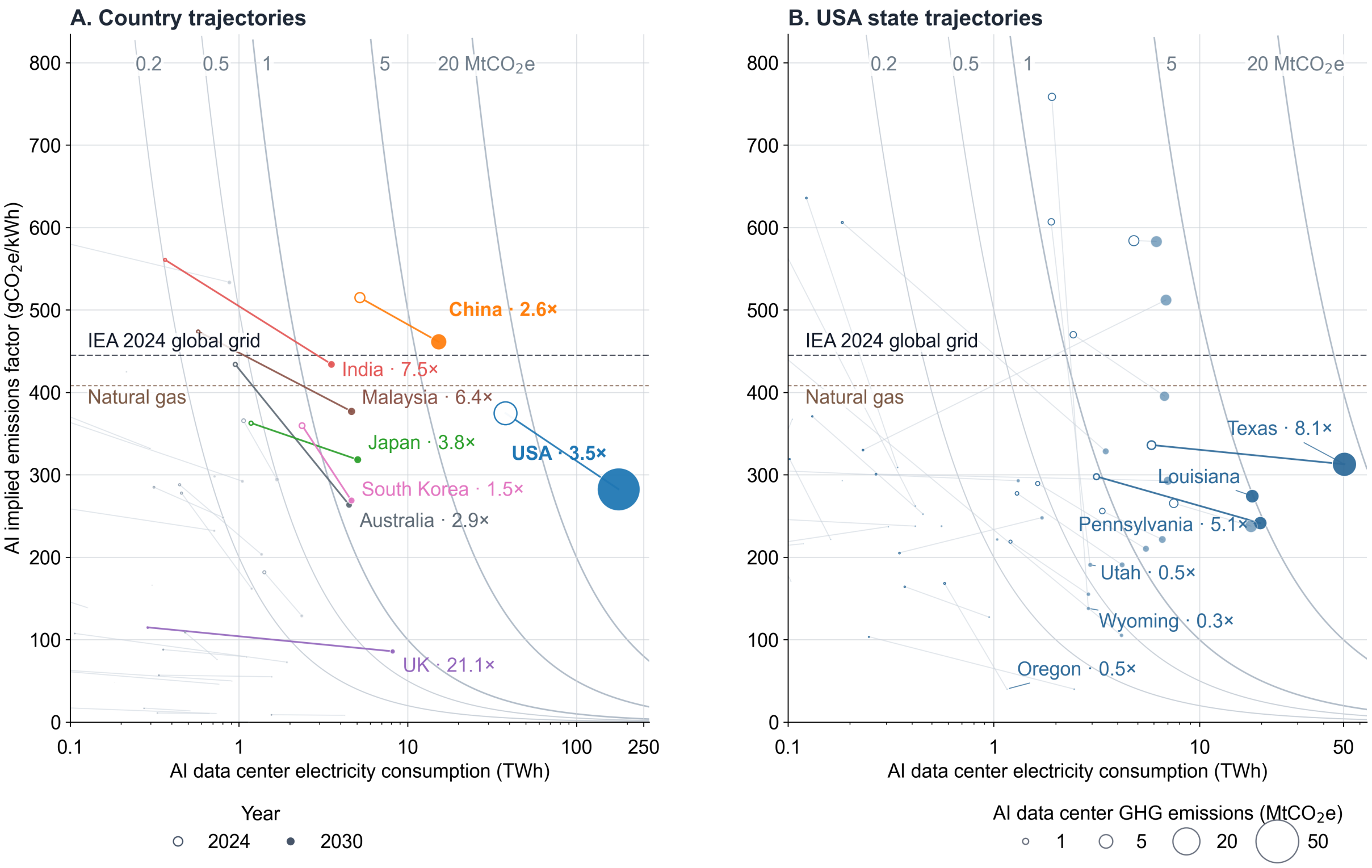

Fig. 3. Country and U.S.-state AI climate trajectories link electricity growth, grid carbon intensity and 2030 emissions under the Current Policies. (A) National trajectories of AI data center electricity consumption (horizontal axis) against the implied AI emissions factor (vertical axis). Open and filled circles mark 2024 and 2030, circle area denotes AI-related GHG emissions. Labels report each highlighted country's 2030/2024 emissions growth ratio; dashed reference lines mark the U.S. eGRID net-generation-weighted direct emission factor for gas-category power plants and the IEA 2024 global grid average[52,53]; Curved lines from top left to bottom right are iso-emission contours representing constant total AI-related GHG emissions. Implied emissions factors used for the vertical axis are the weighted averages including grid electricity and behind-the-meter generation.

Under Current Policies, only 6 out of 80 regions (countries and US states), decrease their absolute GHG emissions inventory of AI-oriented data center operation between 2024 and 2030 (Fig. 3A; Supplementary Fig. 6). The number of regions with absolute reduction in GHG inventory rises to 17 under NDC scenario and to 32 under Net Zero by 2050 scenario (Supplementary Fig. 6).

This race can be visualized as two opposing changes in the dimensions of a rectangle, whose area—the product of its width and height—represents the relative magnitude of GHG emissions (Fig. 4). Grid decarbonization reduces the height of the rectangle by lowering GHG emissions intensity, whereas AI-oriented data-center expansion increases its width by increasing electricity demand. The United States, for example, shows the largest increase in rectangle area because the expansion in width far exceeds the reduction in height. At the same time, grid decarbonization avoids substantial amount of emissions, represented by the greyed-out area above the 2030 rectangle; in the United States, these avoided emissions are larger than the total 2030 emissions of any other country.

Behind-the-meter generation alters this balance by shielding part of the new AI load from grid decarbonization. BTM is projected to supply 20% of U.S. AI-oriented data center electricity by 2030 as grid-interconnection and transmission constraints delay new connections[54,55] (Fig. 4B). The 2024 BTM electricity supply mix, weighted by AI-attributable electricity supplied, consists almost entirely of natural-gas simple-cycle generation (99.97%), with point-of-use hydrogen fuel cells contributing 0.03% (Supplementary Note 11). By 2030, the documented BTM electricity supply mix on the same basis comprises natural-gas simple-cycle turbines (83.08%), hydrogen fuel cells (9.81%), natural-gas reciprocating engines (5.86%) and combined-cycle generation (1.25%) (Supplementary Note 11). BTM generation is already more GHG-intensive than the grid electricity of almost all the host countries in 2024 and remains so through 2030 under all three policy scenarios (Fig. 4). Because BTM generation is largely insulated from local and national grid-decarbonization efforts, it produces additional GHG emissions relative to the all-grid case; we refer to this difference as ‘BTM emissions penalty’. In Fig. 4, this penalty is represented by the red area within the USA-BTM column above the grid-emissions line. In 2024, both BTM use and its emissions penalty are relatively small (Fig. 4A). By 2030, however, BTM becomes a first-order determinant of AI-oriented data-center GHG inventory; under Current Policies, emissions from U.S. facilities using BTM exceed those from all other countries combined (Fig. 4B). Its importance grows further as the grid decarbonizes (Fig. 4C,D). Under the Net Zero 2050 scenario, the U.S. BTM emissions penalty reaches 35% of the global AI-oriented data-center GHG inventory and exceeds the combined AI-data-center emissions of all other countries (Fig. 4D; Supplementary Note 11, Supplementary Table 17 and Supplementary Table 21).

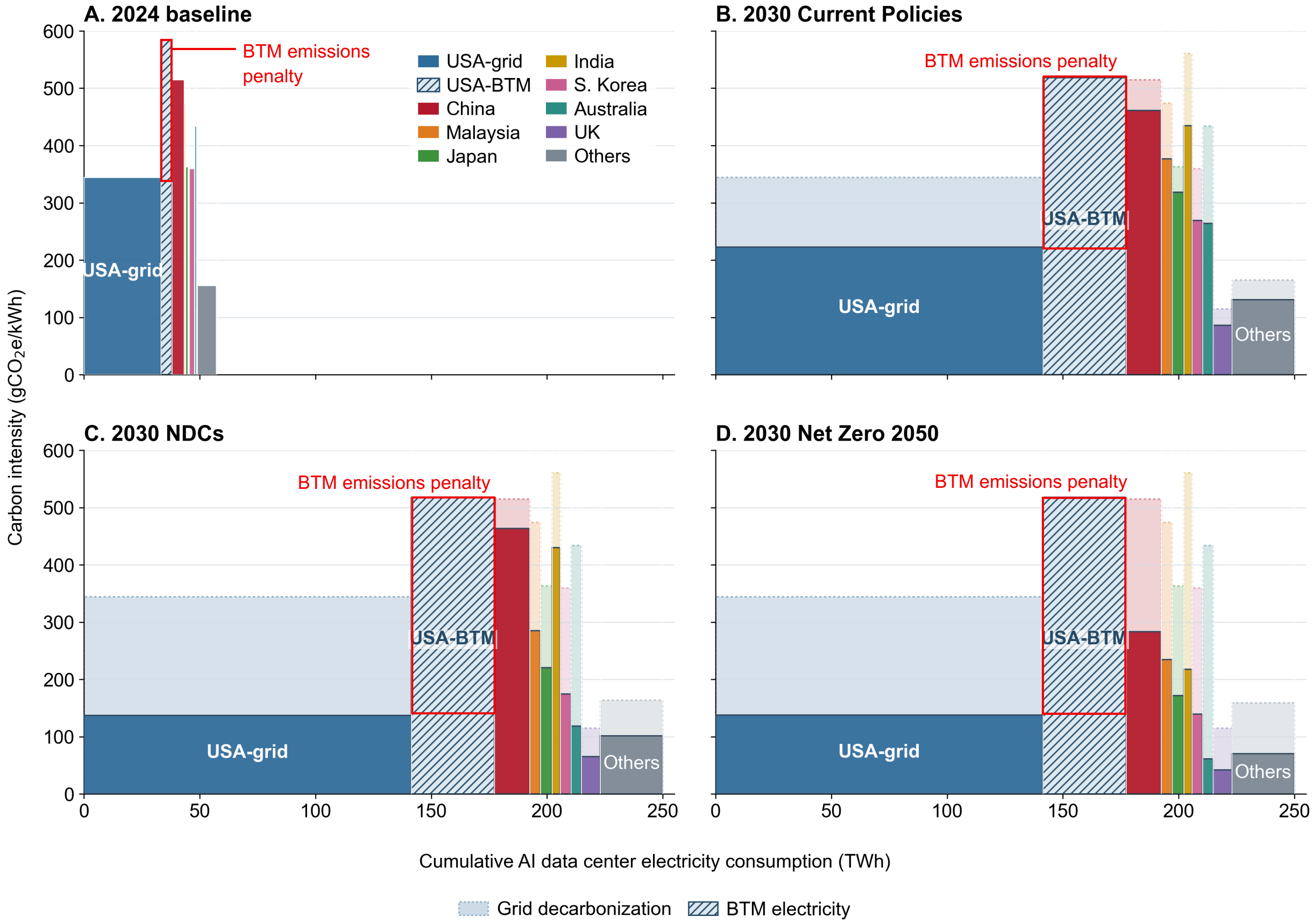

Fig. 4. Cumulative electricity and GHG mechanisms across countries. Countries are ordered by 2030 AI-related GHG emissions along a shared cumulative-capacity axis (in electricity consumption); bar height gives carbon intensity of electricity, so rectangle area equals GHG emissions. (A) The 2024 baseline. (B) The 2030 Current Policies scenario. (C) The 2030 NDC scenario. (D) The 2030 Net Zero by 2050 scenario. These scenarios are drawn from three integrated assessment models including GCAM, MESSAGEix-GLOBIOM and REMIND-MAgPIE. Light extensions show the carbon intensity avoided through grid decarbonization between 2024 and 2030.

## 4 Discussion

This study resolves the climate consequences of AI expansion at the level of the physical infrastructure through which that expansion occurs. By linking a global inventory of AI-oriented data centers to facility locations, projected capacity expansion, host-grid decarbonization trajectories and publicly announced BTM power supply, we show that AI's near-term climate trajectory reflects a structural mismatch between where and how rapidly capacity is being added and how quickly the electricity supplying it is decarbonizing. AI-oriented capacity is concentrated in a relatively small set of power systems, without a corresponding shift towards the fastest-decarbonizing geographies, such that declining grid carbon intensity does not keep pace with projected load growth. Announced BTM generation can further erode this mitigation effect by partially decoupling new AI capacity from improvements in grid carbon intensity. The climate consequences of AI expansion therefore depend jointly on where capacity is built, how rapidly host power systems decarbonize and how new facilities are powered.

Though capacity expansion remains the dominant upward pressure on AI-data-center emissions, aggressive grid decarbonization is shown to be able to materially mitigate AI's climate impact. In the state of Oregon, for example, AI electricity demand is expected to nearly double, but grid emissions intensity is projected to fall by roughly three-quarters under the current policies, allowing absolute emissions to approximately halve. This direction is consistent with Oregon's statutory clean-electricity pathway, which requires an 80% reduction in emissions from covered electricity suppliers by 2030[56].

However, the mitigation potential of aggressive grid decarbonization can be substantially eroded by widespread BTM generation in regions undergoing rapid AI data-center expansion. Texas illustrates this interaction. The state has decarbonized its grid rapidly alongside exceptional renewable deployment, leading the United States in wind generation and reaching nearly 66 GW of utility-scale renewable capacity by 2024, with solar capacity also expanding rapidly in recent years[57]. This trend is consistent with the modelled 41.6% decline in Texas grid emissions intensity by 2030, supported by continued coal displacement and renewable growth.[58] Without this decarbonization, Texas-hosted AI data centers would emit about 17 Mt$CO_2$e in 2030; grid decarbonization alone would reduce this by roughly 7 Mt$CO_2$e, to 10 Mt$CO_2$e. However, documented BTM capacity rises from effectively zero in 2024 to 5.3 GW by 2030, accounting for 52.3% of total capacity additions and increasing projected emissions to 16 Mt$CO_2$e. Thus, BTM generation is projected to erase nearly all of the emissions reductions otherwise achieved through grid decarbonization in Texas.

Grid congestion and multi-year interconnection delays are increasingly constraining the pace at which new data centers can obtain grid power[59]. As AI capacity expands faster than grid infrastructure, natural-gas-based behind-the-meter generation is already emerging as a rapid time-to-power response to these constraints[60,61]. Because this shift directly moves new electricity demand outside the decarbonization trajectory of the host grid, constraining carbon-intensive BTM generation should become an explicit component of data-center and power-sector policy. This requires not only limiting reliance on unabated natural-gas-based BTM generation, but also making rapidly deployable low-carbon alternatives competitive on the same deployment timescale. Faster interconnection and transmission expansion should therefore be coupled with accelerated deployment of clean generation, storage and flexible connection arrangements, while siting incentives should direct new capacity toward less-constrained and faster-decarbonizing power systems. For BTM systems deployed primarily as a time-to-power response, policy should also prioritize a timely transition to grid supply once interconnection becomes available. If lower-carbon grid electricity then supplies routine demand and onsite generation is retained for backup, both the BTM electricity-supply share and its associated emissions penalty would decline. Mandatory disclosure of BTM capacity, fuel use and associated emissions is also needed to prevent this growing source of emissions from remaining outside conventional grid accounting.

## 5 Conclusions

Documented expansion of AI-oriented data centers is projected to raise operational electricity-related GHG emissions through 2030 under all three grid-policy pathways. Facility-level attribution shows how geographic concentration and power-supply configuration determine the offset provided by grid decarbonization. Routine fossil-based behind-the-meter supply weakens that offset. Coordinated siting and grid interconnection, together with access to rapidly deployable low-carbon electricity, can address the supply choices that shape this emerging emissions burden. The inventory provides a traceable basis for assessing those choices while retaining uncertainty in facility identification and planned capacity realization.

## CRediT authorship contribution statement

Xiaoyu Xie: Methodology, Investigation, Formal analysis, Data curation, Visualization, Writing – original draft. John Bistline: Writing – review & editing. Ming Xu: Writing – review & editing. Sangwon Suh: Conceptualization, Methodology, Supervision, Writing – original draft, Writing – review & editing.

## Funding

Sangwon Suh was supported by Xinghua Chair funds. Ming Xu was supported by the National Natural Science Foundation of China (grant no. 52430004). The funders had no role in study design; data collection, analysis or interpretation; manuscript preparation; or the decision to submit the article for publication.

## Declaration of competing interest

The authors declare that they have no known competing financial interests or personal relationships that could have appeared to influence the work reported in this paper.

## Data availability

All processed and intermediate data required to reproduce the reported results are provided in Supplementary Data 1 and the uploaded AI-DC Inventory — Reviewer Reproducibility Package v1.0.0. The package includes the frozen 9,946-row facility inventory and all analysis inputs and manifests. A DOI-minted public archive will be released upon publication.

## Code availability

All custom code required to reproduce the reported analyses, tables and figures is available to editors and reviewers in the uploaded AI-DC Inventory — Reviewer Reproducibility Package v1.0.0. This package provides the complete executable workflow and all required input and intermediate data. A DOI-minted public archive will be released upon publication.

## Declaration of generative AI and AI-assisted technologies in the manuscript preparation process

During the preparation of this work, the authors used LLM-AI for code assistance, language editing, document formatting and reference verification. After using this tool, the authors reviewed and edited the outputs as needed and take full responsibility for the content of the publication.